\documentclass[prl,english,preprintnumbers,amsmath,amssymb,nofootinbib,twocolumn,superscriptaddress]{revtex4-1}
\usepackage[utf8]{inputenc}
\usepackage[T1]{fontenc}
\usepackage{amsmath}
\usepackage{amssymb}
\usepackage{graphicx}
\usepackage{tabularx}
\usepackage{bm}
\usepackage{pifont}
\usepackage{float}

\usepackage{color}
\definecolor{orange}{rgb}{1,0.5,0}

\definecolor{bblue}{rgb}{0.2,0.8,1.0}

\newcommand{\beq}{\begin{equation}}
\newcommand{\eeq}{\end{equation}}
\newcommand{\bea}{\begin{eqnarray}}
\newcommand{\eea}{\end{eqnarray}}

\newcommand{\f}[2]{\frac{#1}{#2}}

\newcommand{\dtcl}{\delta t_{\text{CL}}}

\usepackage[resetlabels,labeled]{multibib}

\begin{document}

\title{Finite-Temperature Equation of State of Polarized Fermions at Unitarity}

\author{Lukas Rammelm\"uller}
\affiliation{Institut f\"ur Kernphysik (Theoriezentrum), Technische Universit\"at Darmstadt, D-64289 Darmstadt, Germany}
\affiliation{GSI Helmholtzzentrum f\"ur Schwerionenforschung GmbH, Planckstra{\ss}e 1, D-64291 Darmstadt, Germany}

\author{Andrew C. Loheac}
\affiliation{Department of Physics and Astronomy, University of North Carolina, Chapel Hill, North Carolina 27599, USA}

\author{Joaqu\'in E. Drut}
\affiliation{Department of Physics and Astronomy, University of North Carolina, Chapel Hill, North Carolina 27599, USA}

\author{Jens Braun}
\affiliation{Institut f\"ur Kernphysik (Theoriezentrum), Technische Universit\"at Darmstadt, D-64289 Darmstadt, Germany}
\affiliation{FAIR, Facility for Antiproton and Ion Research in Europe GmbH, Planckstra{\ss}e 1, D-64291 Darmstadt, Germany}
\affiliation{ExtreMe Matter Institute EMMI, GSI, Planckstra{\ss}e 1, D-64291 Darmstadt, Germany}

\begin{abstract}
We study in a nonperturbative fashion the thermodynamics of a unitary Fermi gas over a wide range of temperatures and spin polarizations. To this end, we use the complex Langevin method, a first principles approach for strongly coupled systems. Specifically, we show results for the density equation of state, the magnetization, and the magnetic susceptibility. At zero polarization, our results agree well with state-of-the art results for the density equation of state and with experimental data. At finite polarization and low fugacity, our results are in excellent agreement with the third-order virial expansion. In the fully quantum mechanical regime close to the balanced limit, the critical temperature for superfluidity appears to depend only weakly on the spin polarization.
\end{abstract}

\maketitle

{\it Introduction --}
Without a doubt, one of the most intensely studied systems in recent years, at the interface of atomic, nuclear, and high-energy physics, is that of two-component fermions in the scale-invariant limit of infinite $s$-wave scattering length and effectively zero interaction range: the unitary Fermi gas (UFG)~\cite{Bertsch, Kaplan:1998tg, Schafer:2009dj, ZwergerBook}. This system is now routinely realized to an excellent approximation with ultracold alkali atoms in several laboratories around the world (see Refs.~\cite{Inguscio:2007cma, Giorgini2008, Bloch2008, RevModPhys.82.1225, ZwierleinReview} for reviews of theory and experiment) and simultaneously (though only approximately) in dilute neutron matter in neutron star crusts~\cite{Schwenk:2005ka, Kolomeitsev:2016sjl, Strinati:2018wdg}. Because of the lack of scales characterizing the interaction between the fermions, all physical quantities at unitarity are fully determined by {\it universal} numbers in units of the fermion density~\cite{Universality1}, that being the only scale of the system. This property renders the system relevant for such disparate energy scales as those of atomic and astrophysics and has, moreover, been shown to reflect a nonrelativistic type of conformal invariance~\cite{MSW, PhysRevD.76.086004, PhysRevD.85.106001, PhysRevA.86.013616}.

A peculiarity of the UFG is that it lies in the middle of the crossover between Bardeen-Cooper-Schrieffer superfluidity and Bose-Einstein condensation where the appearance of pseudogap phenomena and preformed Cooper pairs at high temperature appears possible~\cite{PhysRevB.66.024510, CHEN20051, PhysRevLett.103.210403, PhysRevLett.107.145304, PhysRevLett.110.090401}. This suggests intriguing connections to high-$T_c$ superconductors. Because of such relevance of the UFG for various fields, the past two decades have seen uncounted studies exploring the properties of this crossover in the unpolarized limit both theoretically and experimentally~\cite{ZwergerBook}. Finite spin polarizations are even more challenging to tackle (see e.g.~\cite{RevModPhys.76.263, RevModPhys.86.509, 2010RPPh...73k2401C, GUBBELS2013255, 0034-4885-81-4-046401} for reviews, and~\cite{Zwierlein492, Nature44254, PhysRevLett.97.030401, Partridge503, PhysRevLett.97.190407, Nature451689, PhysRevLett.103.170402, 2009PhRvL.102w0402S,Nature4631057, PhysRevLett.106.215303, Nature472201, NJPhys13, PhysRevLett.107.145305, PhysRevA.92.063616, PhysRevLett.117.235301} for experimental work) and therefore this case leaves us with many puzzles. At low temperatures, when the system is superfluid, a large enough polarization will destroy superfluidity~\cite{Chandra,Clogston}. Precisely how that happens, and what other exotic superfluid phases may be traversed in the process, has remained a controversial topic not only for atomic superfluids but also for their quantum chromodynamics (QCD) counterparts, namely color superconductors~\cite{RevModPhys.76.263}. Part of the challenge in answering such questions is that the UFG (not unlike QCD and many other systems) is a strongly correlated many-body system lacking a small parameter and therefore can only be tackled with nonperturbative methods. However, nonperturbative (semi-)analytic studies of such systems rely on some ansatz and conventional Monte Carlo (MC) calculations are unavailable at finite polarization due the infamous sign problem.

In this Letter, we explore the {\it spin polarized} UFG at finite temperature, providing some of the essential measurable properties that characterize its universal thermodynamics, namely, the density and magnetic equation of state (EOS). From those, differentiation yields static response functions such as the compressibility and magnetic susceptibility, while integration yields the pressure. To determine those equations of state, we implement a complex version of stochastic quantization known as the complex Langevin (CL) method~\cite{Parisi:1984cs}, which we have developed and tested for spin- and mass-imbalanced one-dimensional nonrelativistic systems~\cite{Loheac:2017yar}, including successful comparisons with exact answers in the ground state~\cite{Rammelmuller:2017vqn} and at finite temperature~\cite{Loheac:2018yjh}. In the present Letter, we further validate our approach by comparing our results with the virial expansion and state-of-the-art MC calculations at zero polarization, eventually obtaining {\it ab initio} predictions for thermodynamic quantities of the UFG over wide temperature and polarization ranges.

{\it Hamiltonian and method --}
Fermions in the unitary limit are governed by a Hamiltonian with a nonrelativistic dispersion relation and a zero-range interaction,
\begin{align}
\hat H \!=\! {\int{\!d^3 x\,\hat{\psi}^{\dagger}_{s}({\bf x})\left(\!-\frac{\hbar^2\nabla^2}{2m}\!\right)\hat{\psi}_{s}({\bf x})}}\!-\!
g\! \int{\!d^3 x\,\hat{n}_{\uparrow}({\bf x})\hat{n}_{\downarrow}({\bf x})}\,,\nonumber
\end{align}
where $\hat{\psi}^{\dagger}_{s}, \hat{\psi}^{}_{s}$ are the fermion creation and annihilation operators, respectively, for spin projection $s=\uparrow,\downarrow$ (summed over in the kinetic term), and the corresponding coordinate-space densities are $\hat{n}_{s}({\bf x})=\hat{\psi}^{\dagger}_{s}({\bf x})\hat{\psi}_{s}({\bf x})$. Although we have written $\hbar$ and the fermion mass $m$ explicitly, we {take $\hbar = k_\text{B} = m = 1$ from this point on. The grand-canonical partition function then reads
\beq
\label{Eq:PartitionZDef}
\mathcal Z = \text{Tr} \exp \left[{-\beta( \hat H - \mu_\uparrow \hat N_\uparrow - \mu_\downarrow \hat N_\downarrow)}\right]\,,
\eeq
where $\mu_s$ is the chemical potential for spin $s=\uparrow,\downarrow$ particles, $\hat N_s$ is the corresponding particle number operator, and $\beta^{-1}=T$ is the temperature. To study the strongly coupled many-body problem described by~$\mathcal Z$, we put the system on a spacetime lattice (via a Suzuki-Trotter factorization) and introduce a path integral representation of the interaction by way of an auxiliary-field Hubbard-Stratonovich (HS) transformation. As those steps are rather standard (see, e.g.,~\cite{Drut:2012md}), we only state the result,
\beq
\label{Eq:PartitionZPathIntegral}
\mathcal Z = \int \mathcal D \sigma e^{-S[\sigma]}\,,
\eeq
where $S[\sigma] = -\ln \det ( M_\uparrow[\sigma] M_\downarrow[\sigma])$ is the action for the (real-valued) HS field $\sigma$ and contains all the input parameters mentioned above. The details of the shape of the real-valued Fermi matrix $M_s[\sigma]$ can be found, for instance, in Ref.~\cite{Drut:2012md}. It is important to note here, however, that $M_\uparrow[\sigma]$ includes $\mu_\uparrow$ and not $\mu_\downarrow$, and vice versa for $M_\downarrow[\sigma]$; i.e., we use a HS transformation that decouples the interaction in the density channel.
As a result, in the unpolarized limit $\mu_\uparrow = \mu_\downarrow$, the fermion determinant is positive and the action is real, such that $e^{-S[\sigma]} \geq 0$ can be used as a probability measure in a Metropolis-based MC calculation, i.e. there is no sign problem in that case. On the other hand, for the polarized case $\mu_\uparrow \neq \mu_\downarrow$, such that $M_\uparrow[\sigma] \neq M_\downarrow[\sigma]$, and therefore $S$ can be complex, {which hinders the use of} probabilistic MC approaches.

The aforementioned sign problem is well known and pervades MC approaches across all of physics~\cite{Troyer:2004ge}, including high-$T_c$ superconductors (due to strong repulsive interaction away from half filling)~\cite{PhysRevX.5.041041}, nuclear structure (strong repulsive core, finite spin-isospin polarization)~\cite{MeissnerFewBody, Epelbaum34612_2}, and QCD (at finite quark density)~\cite{Lombardo:2006yc, deForcrand:2010ys, Philipsen:2012nu, Gattringer:2016kco}, to name a few. Recently, some progress has been made in understanding the sign problem as well as in its treatment with complex-plane methods such as the CL approach~\cite{Parisi:1984cs} and Lefschetz thimbles~\cite{Cristoforetti:2012su, Cristoforetti:2013wha, Fujii:2013sra, Aarts:2014nxa, Alexandru:2015sua}. In essence, the CL algorithm implements an extension of conventional, Langevin-based stochastic quantization~\cite{Parisi:1980ys, Damgaard:1987rr, Batrouni:1985jn} to the case of complex-valued actions. As the Langevin equation uses $S$ to evolve $\sigma$ in its configuration space, a complex $S$ naturally requires complexifying the HS field $\sigma$. Further details on the algorithm and our implementation can be found in Refs.~\cite{Loheac:2017yar, Loheac:2018yjh, Drut:2017fsv, Loheac:2017uuj, Rammelmuller:2017myk}. Thus far, we have successfully applied such an approach to nonrelativistic fermions in 1D in a variety of situations, such as finite temperature and polarization~\cite{Loheac:2018yjh}, and mass asymmetry at zero temperature~\cite{Rammelmuller:2017vqn}. Those studies yielded an optimistic outlook for their higher-dimensional counterparts, i.e., this Letter. Still, a word of caution is in order regarding this method. While conventional Metropolis-based methods are on solid mathematical footing at vanishing polarization, the CL approach remains a method under construction. {A discussion} of the issues is beyond the scope of this Letter, but these are being investigated by us and other groups in the lattice QCD {area (see, e.g.,~\cite{Aarts:2009uq, Aarts:2011ax, Aarts:2017vrv, Nagata:2016vkn, Bloch:2017sex}).} We emphasize that the calculations presented below display the same run-time features as our prior 1D studies which, together with the self-consistency of the results and the agreement with other methods in the balanced case and the virial expansion at finite polarization, gives some confidence on the reliability
of the answers.

{\it Results --}
To characterize the universal thermodynamics of the polarized UFG, we computed the density $n$, magnetization $m$, and normalized magnetic susceptibility~$\bar{\chi}^{}_M = \partial \bar{m}/\partial (\beta h)$ with~$\bar{m}=m/n(\beta h\!=\! 0)$ as functions of the dimensionless chemical potential $\beta\mu = \beta(\mu_\uparrow + \mu_\downarrow)/2$, and the dimensionless chemical potential difference $\beta h = \beta (\mu_\uparrow - \mu_\downarrow)/2$. The path integral form of the thermal expectation values of $n$ and $m$ is obtained by differentiating $\ln \mathcal Z$ with respect to $\mu$ and $h$. The magnetic susceptibility, which becomes the Pauli susceptibility in the noninteracting case, is then derived from the magnetic EOS. To evaluate such path integrals, we discretized spacetime into a (3+1)-dimensional lattice of spatial volume $V = L^3$, with $L = \ell N_x$, $N_x = 7,\,9,\, 11$, lattice spacing $\ell = 1$, and periodic boundary conditions. For the temporal direction, we chose $N_\tau = 160$, with temporal lattice spacing $\tau = 0.05 \ell^2$, and antiperiodic boundary conditions for the fermion fields. Note that, while we varied the spatial extent of the box in our calculations, we kept~$\beta = \tau N_\tau = 8.0$ fixed. Our choice for the latter determines the thermal de Broglie wavelength $\lambda_T = \sqrt{2\pi \beta}\simeq 7.0$ being consistent with the continuum-limit window $1 = \ell \ll (\lambda_T, \lambda_F) \ll L = N_x \ell$, where $\lambda_F=2\pi/k_F$ is the Fermi wavelength, and  $k_F = (3\pi^2 n)^{\frac{1}{3}}$ is the Fermi momentum at the given density. Thus, the computational challenge, besides the sign problem, is that of opening that window of scales by making $N_x$ and $\beta$ as large as possible, in that order, and staying in a dilute regime to suppress artifacts associated with the ultraviolet energy cutoff imposed by the lattice. {Note that the reliability of calculations based on our present set of spacetime lattice parameters has been analyzed in detail in the past~\cite{PhysRevLett.96.160402,LEE2009117,PhysRevA.86.013604,PhysRevA.87.023615,Bulgac2008,PhysRevLett.106.205302}.}

The bare coupling constant $g$ in the Hamilton operator was fixed to the two-body bound-state threshold using L\"uscher's formula~\cite{Luescher1, Luescher2}, as in Ref.~\cite{Drut2012}. Under those conditions, we varied the asymmetry parameter over the range $\beta h \in [0.0,2.0]$ (corresponding to $T \geq h/2$), and the chemical potential in the interval $\beta {\mu} \in [-3.0, 2.5]$, covering the semiclassical regime (at low fugacities $z_s=e^{\beta\mu_s}$, where the virial expansion is valid) to the fully quantum mechanical regime at large positive $\beta\mu$, including a small region below the superfluid transition temperature for the unpolarized system, at $(\beta \mu)_c \simeq 2.5$~\cite{Ku2012, Boettcher:2014tfa, Roscher:2015xha, Goulko:2015rsa}.
\begin{figure}
  \centering
  \includegraphics[width=\columnwidth]{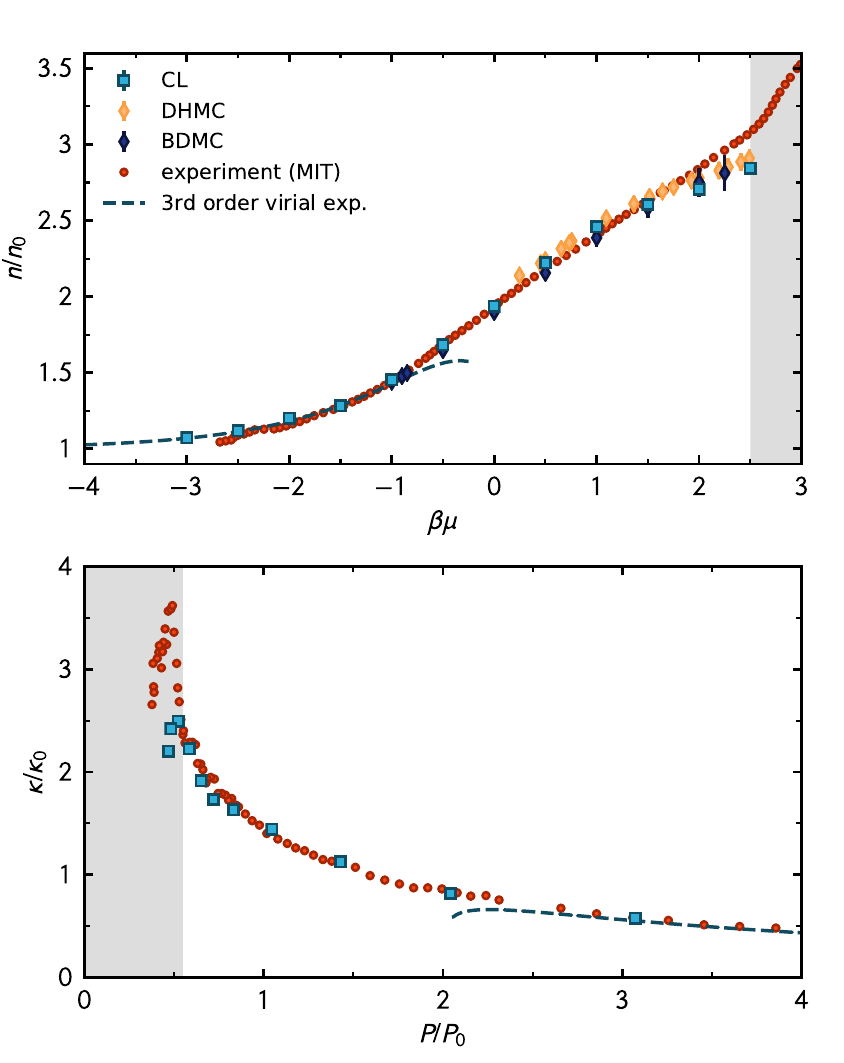}
  \caption{\label{Fig:EOSbalanced}
  (Top) Density of the balanced UFG obtained by CL (blue squares), in units of the noninteracting
  unpolarized density ${n}_0$ as a function of the dimensionless average chemical potential $\beta\mu$. Also shown, third-order virial expansion (dashed line), experimental results of Refs.~\cite{Ku2012, VanHoucke2012} (red circles), and theoretical results obtained by bold diagrammatic Monte Carlo calculations ~\cite{VanHoucke2012} (dark diamonds) and determinantal hybrid Monte Carlo calculations \cite{Drut2012} (light diamonds).
  (Bottom) Compressibility $\kappa$ as derived from the density EOS (see Supplemental Material \cite{suppl_material}) in units of its noninteracting  ground-state value $\kappa_0$, as a function of the pressure $P$ normalized by the noninteracting ground-state pressure $P_0$ (blue squares), compared to experimental values \cite{Ku2012} (red circles) and third-order virial expansion (dashed line). Statistical uncertainties for the CL results are on the order of the symbol sizes. Shaded areas indicate the superfluid phase.}
\end{figure}

To validate our results, we use prior lattice MC results~\cite{Drut2012}, diagrammatic MC results~\cite{VanHoucke2012} and MIT experimental~\cite{Ku2012, VanHoucke2012} results obtained in the unpolarized limit {(first measured in~\cite{OHara, PhysRevLett.98.080402} and computed with MC calculations in~\cite{Bulgac2006})}, as well as the third-order virial expansion at finite polarization, which reads
\bea
n - n_0 &=& \frac{Q_1}{V} \left [ 2 \Delta b_2  z_\uparrow z_\downarrow +  3\frac{\Delta b_3}{2} (z^2_\uparrow z_\downarrow+ z_\uparrow z^2_\downarrow) \right ]\, ,
\eea
\bea
m - m_0 &=& \frac{Q_1}{V} \left [ \Delta b_3 (z^2_\uparrow z_\downarrow - z_\uparrow z^2_\downarrow) \right ],
\eea

where $Q_1$ is the two-species single-particle partition function, $V$ is the spatial volume, and in the continuum $Q_1/V \to 2/\lambda_T^3$. The interacting total density is given by~$n$, $n_0=n_0(\beta \mu, \beta h)$ is the noninteracting total density, $m=n_{\uparrow}-n_{\downarrow}$ is the magnetization of the interacting system, and~$m_0=m_0(\beta\mu, \beta h)$ is the associated noninteracting magnetization.
The above coefficients are $\Delta b_j = b_j - b_j^{0}$, where $b_j^{0} = (-1)^{j-1} j^{-5/2}$ are the virial coefficients of the noninteracting gas, and $b_2 = 3/(4 \sqrt{2})$ and $b_3 \approx -0.29095$ {(see, e.g., Refs.~\cite{Nature4631057,Liu2012}) are} the coefficients of the unitary gas. The coefficient $b_4$ is also known for the unpolarized gas: $b_4=0.078(18)$ (see Ref.~\cite{Yan2016}), but two separate coefficients are needed at that order in the polarized case.

\begin{figure*}
  \centering
  \includegraphics[width=2.\columnwidth]{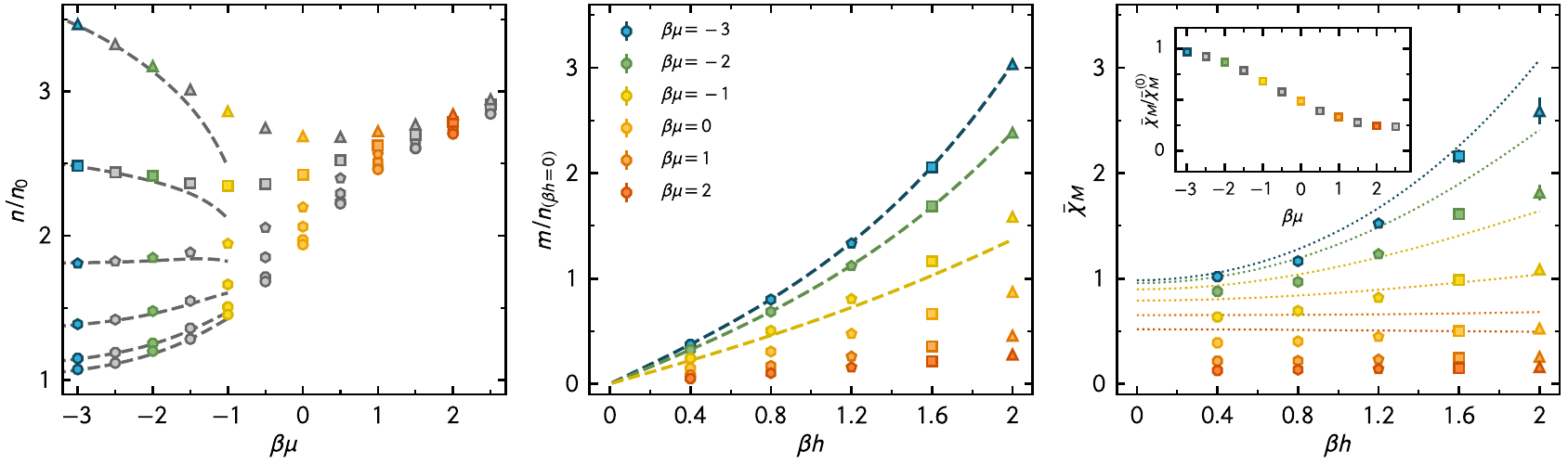}
  \caption{\label{Fig:EOSimbalanced}
  (Left) Density of the UFG in units of the noninteracting density from bottom to top:~$\beta h = 0$ (circles), $0.4$ (octagons), $0.8$ (hexagons), $1.2$ (pentagons), $1.6$ (squares), $2.0$ (triangles), compared to the third-order virial expansion (dashed lines). Colors encode fixed values of $\beta \mu$ shown in all panels.
  (Center) Magnetization in units of the interacting density for the balanced system as a function of $\beta h$ for several values of $\beta\mu$. For~$\beta\mu \leq -1.0$, third-order virial expansion is shown with dashed lines.
  (Right) Dimensionless magnetic {susceptibility~$\bar{\chi}^{}_M$ as a function of $\beta h$ (symbols) compared} to the corresponding susceptibility of the free Fermi gas~$\bar{\chi}^{0}_M$ (dotted lines) at equal chemical potential and asymmetry (color and shape coding as in other panels).
  (Inset) Ratio~${\bar \chi}^{}_M / {\bar \chi}^{0}_M$ as a function of~$\beta \mu$ at $\beta h = 0.4$.}
\end{figure*}

For the parameter region studied, we find that our $\beta h = 0$ results are in excellent agreement with the third-order virial expansion for~$\beta \mu \lesssim -1$; see Fig.~\ref{Fig:EOSbalanced} for the density EOS and the isothermal compressibility~$\kappa = (1/n)(\partial n/\partial P)|_T$ with~$P$ being the pressure and~$n$ being the total density. Moreover, our results reproduce closely the existing results from lattice MC~\cite{Drut2012}, diagrammatic MC~\cite{VanHoucke2012}, and the MIT experiments~\cite{Ku2012, VanHoucke2012} in the unpolarized limit, up to $\beta \mu = 2.0$, which reflects the smallness of the systematic effects in that parameter range. The smoothness of the curve connecting the data points shows that statistical effects are also well controlled and are roughly of the size of the symbols. For $\beta \mu > 2.0$, on the other hand, systematic effects in all state-of-the-art calculations, namely finite-range and finite-volume effects, become more important and underlie the observed deviation from the MIT measurements at low temperature, i.e. close to and below the superfluid phase transition. Still, some indication of the appearance of the phase transition is visible in our present data as a sharp peak in the compressibility close to~$P/P_0 \approx 0.5$, in accordance with experiment.

Given the excellent agreement of our results for the balanced UFG with existing theoretical and experimental data above the superfluid phase transition, we now proceed to the polarized case. In Fig.~\ref{Fig:EOSimbalanced}, we present our main results: density EOS normalized by the density of the noninteracting {gas $n_0(\beta\mu, \beta h\! =\! 0)$ as} a function of~$\beta\mu$ (left panel) for $\beta h = 0, 0.4, \dots, 2.0$; magnetization (central panel) normalized by the interacting density of the {\it balanced} system $n(\beta\mu, \beta h \!=\! 0)$, as well as magnetic susceptibility (right panel) as a function of the asymmetry parameter~$\beta h$ for $\beta \mu = -3, -2, \dots, 2$.

For the density and magnetization equations of state, we again find excellent agreement with the virial expansion for sufficiently negative~$\beta\mu$. However, we also observe that the regime of validity of the expansion appears to shrink as~$\beta h$ is increased, see left panel of Fig.~\ref{Fig:EOSimbalanced}. Indeed, for~$\beta h = 2.0$, the third-order virial expansion clearly deviates from our nonperturbative results for~$\beta \mu \gtrsim -1$, as opposed to the balanced case discussed above.

As~$\beta \mu$ is increased, the equations of state obtained for different values of~$\beta h$ approach the EOS of the balanced system. This is not unexpected, as the relative asymmetry~$h/\mu$ decreases when~$\beta \mu$ is increased at fixed~$\beta h$. Of course, the approach to the balanced EOS should happen at progressively larger values of~$\beta \mu$ when~$\beta h$ is increased, which is indeed the case and can be seen in the left panel of Fig.~\ref{Fig:EOSimbalanced}. As the balanced system is known to be governed by a superfluid ground state above a critical value of~$\beta \mu$, this observation also suggests that the critical temperature decreases with increasing spin asymmetry, in line with (semi-)analytic studies~\cite{2010RPPh...73k2401C, GUBBELS2013255, Boettcher:2014tfa, Roscher:2015xha,Gubbels:2007xc, 2018arXiv180403035F} and lattice MC studies of a slightly spin-imbalanced UFG using reweighting techniques~\cite{2010PhRvA..82e3621G}.

Our discussion of the density EOS at finite spin asymmetry carries over to the magnetization $m$ (Fig.~\ref{Fig:EOSimbalanced}, center). Similar to the density, the results for $m$ match the third-order virial expansion for large negative values of $\beta\mu$. As~$\beta \mu$ is increased, however, our nonperturbative results clearly start to deviate from the virial expansion. For~$\beta \mu = 2.0$, i.e. close to the critical value of the balanced system, we observe that $m$ only shows a very mild dependence on~$\beta h$. As $m$ is expected to be small in the superfluid phase (the response to $h$ being suppressed by the pairing gap; see e.g.~\cite{PhysRevA.74.013614}), our results suggest that the system remains close to the superfluid phase for~$\beta h \lesssim 2$, provided that~$\beta\mu$ is fixed close to its critical value $(\beta \mu)_c \simeq 2.5$ for the balanced case. Sufficiently below $(\beta \mu)_c$, i.e. at sufficiently high temperature, the system can easily ``magnetize'' by increasing $\beta h$.

To supplement our discussion of magnetic properties of the UFG, we also show results for the magnetic susceptibility $\bar{\chi}^{}_M$, which measures the response under a variation of the spin asymmetry (Fig.~\ref{Fig:EOSimbalanced}, right panel). In the noninteracting gas at low effective magnetic field~$\beta h$, the susceptibility is well approximated by the field-independent Pauli susceptibility. For negative~$\beta \mu$, corresponding to the very dilute limit, our results for $\bar{\chi}^{}_M$ of the UFG approach those for the free Fermi gas. Interestingly, even for~$\beta \mu$ close the critical point, the functional form of the susceptibility of the interacting system is still very similar to that of the free Fermi gas, albeit rescaled by a $\beta \mu$-dependent factor. The latter is shown in the inset in the right panel of Fig.~\ref{Fig:EOSimbalanced} at $\beta h = 0.4$.

Let us finally comment on the dependence of the superfluid critical temperature $T_c$ on~$\beta h$. As mentioned above, all of our results display a rather mild dependence on~$\beta h$ for~$\beta \mu \gtrsim 2.0$, which suggests a rather mild dependence of $T_c$ as well, at least in the range $0 \leq \beta h \sim 2.0$. This observation is also supported by a computation of the compressibility: as we increase~$\beta h$, we only observe a very slight shift of the maximum to lower temperatures compared to the balanced case (see Supplemental Material \cite{suppl_material}). This shift appears to be smaller than in \mbox{(semi-)}analytic studies~\cite{2010RPPh...73k2401C, GUBBELS2013255, Boettcher:2014tfa, Roscher:2015xha,Gubbels:2007xc, 2018arXiv180403035F}. However, further work is needed to resolve this dependence quantitatively.

{\it Summary and conclusions --}
We carried out a nonperturbative characterization of the density and magnetization EOS of the UFG at finite temperature. To that end, we implemented a finite-temperature stochastic lattice approach that addresses the sign problem by going to the complex plane; i.e., we used the complex Langevin approach and presented our results as a function of $\beta\mu$ and $\beta h$. We emphasize that those results are experimentally testable predictions~\cite{PhysRevLett.118.123401} for universal properties of quantum many-body physics in the unitary limit, as realized, in particular, with ultracold gases. In the unpolarized case, we recover state-of-the-art results. At finite polarization, our answers agree with the third-order virial expansion for $\beta\mu \lesssim -2.0$, where the expansion is expected to be valid. As in our 1D studies~\cite{Loheac:2018yjh}, however, the expansion deteriorates as $\beta h$ is increased. For increasing~$\beta \mu$, we find that the density EOS at finite asymmetry approaches the EOS of the balanced system.
That approach is ``delayed'' when~$\beta h$ is increased, suggesting a decrease of the critical temperature associated with the superfluid phase transition; this is as expected since $h$ tends to facilitate Cooper pair breaking. Our results for the magnetization support this interpretation and suggest a mild~$\beta h$ dependence even up to $\beta h = 2.0$. The present Letter does not only set the stage for future {\it ab initio} studies of this dependence but also of key features in the low-temperature regime, such as phase separation associated with the Chandrasekhar-Clogston limit, which has already attracted tremendous attention for many years now, both from the experimental~\cite{Zwierlein492, Nature44254, PhysRevLett.97.030401,Nature4631057,PhysRevLett.106.215303,PhysRevA.88.063614} and theoretical side (see, e.g., Refs.~\cite{PhysRevLett.91.247002,PhysRevLett.97.200403,PhysRevA.73.051602,PhysRevA.74.011602,PhysRevA.74.063628,PhysRevA.75.031605,PhysRevA.79.043622,Boettcher:2014tfa}).

\acknowledgments
The authors would like to thank R.~Hulet, J. E. Thomas, J. H. Thywissen, and M. W. Zwierlein for comments on the manuscript. J.B. acknowledges support by the DFG under grant BR 4005/4-1 (Heisenberg program). J.B. and L.R. acknowledge support by HIC for FAIR within the LOEWE program of the State of Hesse. This material is based upon work supported by the National Science Foundation Graduate Research Fellowship Program under Grant No. DGE{1144081}, National Science Foundation Computational Physics Program under Grant No. PHY{1452635}. Numerical calculations have been performed at the LOEWE-CSC Frankfurt.

\bibliography{polarized_ufg_notes}


\clearpage
\renewcommand\thefigure{S\arabic{figure}}
\setcounter{figure}{0}
\setcounter{equation}{0}
\renewcommand{\theequation}{S\arabic{equation}}

\makeatletter
\renewcommand\@biblabel[1]{[S#1]}
\makeatother

\section{Supplemental Material}
 We present a systematic order-by-order comparison of the virial expansion with the density equation of state as obtained from our Complex Langevin calculation. Additionally, we show the compressibility, relative magnetization, and pressure at finite polarization. We discuss the finite-volume effects in the density, Langevin step-size effects, and provide details on how interpolations were performed to obtain derived and integrated quantities. We close by demonstrating that our results satisfy the corresponding Maxwell relation, highlighting the level of self-consistency of our results.

\subsection{Virial expansion}
As it was already discussed in the main text, the virial expansion (VE) provides an important and reliable tool to obtain thermodynamic quantities at temperatures far away from {the fully quantum mechanical regime, i.e. at} small fugacities $z = e^{\beta\mu}$. Since in this regime the fugacity constitutes a small parameter, an expansion can be performed and increasing the expansion order successively improves the range of validity. A natural boundary for the validity of the VE is at $\beta\mu = 0$, i.e. $z = 1$ since then the fugacity is {no longer a small parameter}. For spin-polarized systems, {we have to deal with an expansion in two fugacities (one for each spin flavor). The range of convergence measured in terms of the mean chemical potential~$\mu$ of the two species is then expected to decrease because of the difference in the chemical potentials associated with the two species. Note that the chemical potential of one of the species becomes smaller when the so-called Zeeman field~$h$ is increased.}

In Fig.~\ref{Fig:ve}, we compare the results for the density equation of state from the VE with those obtained from our CL study. Note that, at leading order, the result {from the VE is simply the density} of the non-interacting Fermi gas. We indeed observe that, with increasing order, the results from the VE successively approach those from our non-perturbative study, see Fig.~\ref{Fig:ve}. In accordance with our discussion above, we also find that the range of validity of the VE expansion (in terms of~$\beta\mu$) appears to be maximal for~$\beta h=0$, see Ref.~\cite{Loheac:2018yjh} for a corresponding detailed discussion for one-dimensional systems.
\begin{figure}[t]
  \centering
  \includegraphics[width=\columnwidth]{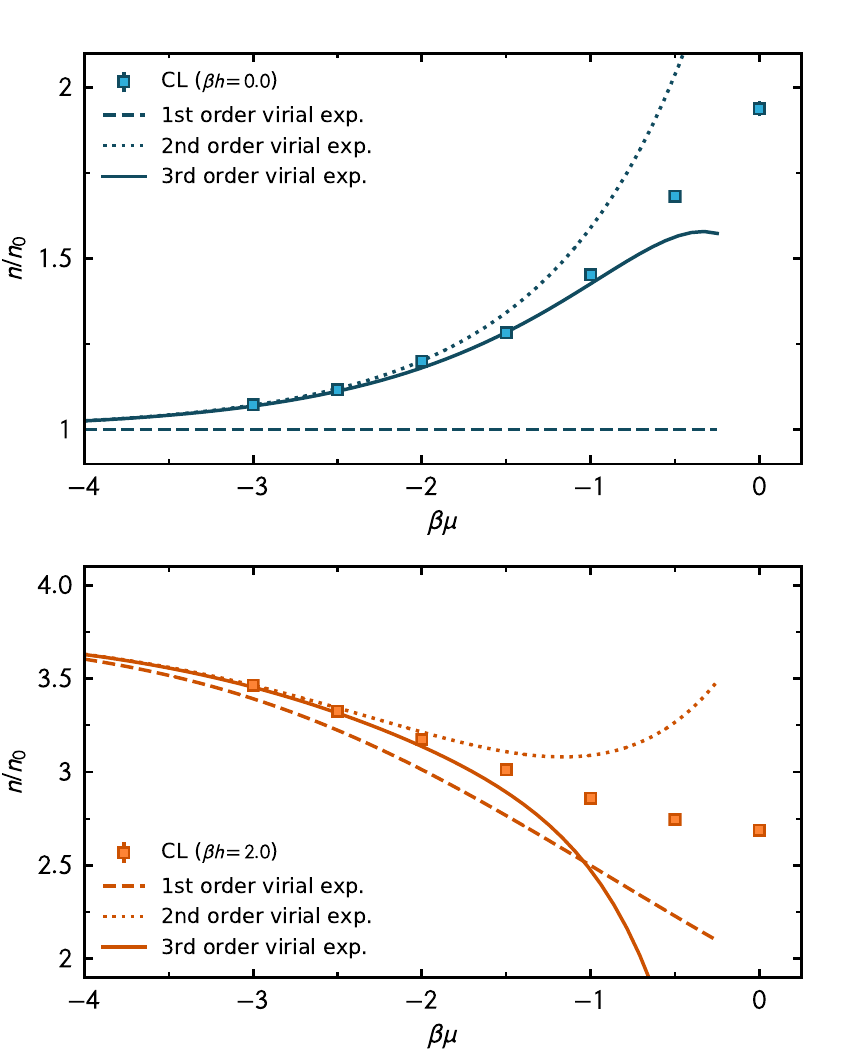}
  \caption{\label{Fig:ve} Virial expansion at first, second and third order compared to the density equation of state as obtained from our CL study (dashed line, dotted line, solid line and symbols, respectively). The top panel shows the spin-balanced case ($\beta h = 0)$ {whereas the bottom panel} shows a spin-polarized system at $\beta h = 2$. }
\end{figure}
\subsection{Compressibility \& relative polarization}
\begin{figure}[t]
  \centering
  \includegraphics[width=\columnwidth]{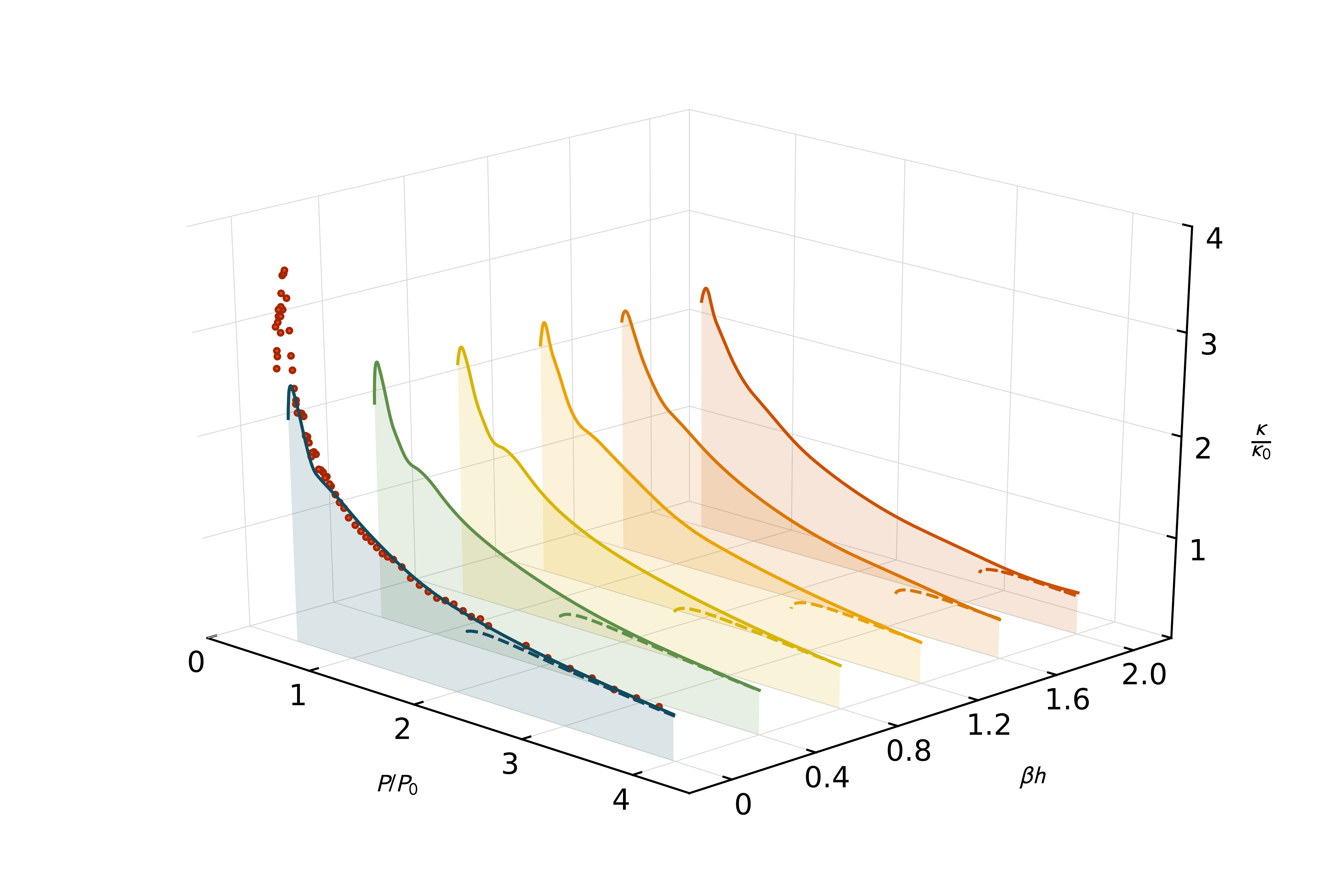}
  \caption{\label{Fig:compressibilities} {Isothermal compressibility for different} values of the chemical-potential asymmetry $\beta h = 0.0, 0.4, 0.8, 1.2, 1.6, 2.0$ (solid lines). Red circles represent experimental data for the balanced gas~\cite{Ku2012} and dashed lines represent the results from the virial expansion at third order.}
\end{figure}
The (isothermal) compressibility~$\kappa$ can be derived directly from the density,
\beq
\label{Eq:kappa}
  \kappa = \frac{1}{n}\left . \frac{\partial n}{\partial P}\right |_T = \frac{\beta}{n^2}\left . \frac{\partial n}{\partial (\beta \mu)}\right |_T\,,
\eeq
and was determined from our MC data by differentiating the density with respect to $\beta \mu$. Note that, in the {bottom panel of Fig.~\ref{Fig:compressibilities} in the main text, the} noninteracting ground-state values are used as scales for the pressure and compressibility at the given density, respectively: $P_0 = 2 n \epsilon_F / 5$ and $\kappa_0 = 3/(2 n\epsilon_F)$, where $\epsilon_F = k_F^2/(2m)$ and $k_F = (3\pi^2 n)^{1/3}$.

To supplement the results shown in the main text, we show the compressibility for various values of the chemical potential asymmetry~$\beta h$ in Fig.~\ref{Fig:compressibilities}. As noted, we observe that the position of the maximum depends only weakly on~$\beta h$, suggesting an almost constant critical temperature (within our accuracy) as a function of~$\beta h$, at least within the considered parameter range~$\beta h \leq 2.0$.

To suppress numerical {artifacts} in the differentiation of the {density EOS with respect to~$\beta \mu$ [Eq.~(\ref{Eq:kappa})]}, we interpolated our data and performed an exact derivative of the interpolating function. Details of this procedure will be discussed below.

In addition to the compressibility, the relative polarization {(or magnetization) is of interest, i.e. the} magnetization of the system relative to the (interacting) density. In Fig.~\ref{Fig:RelativePol}, we show this quantity as a function of~$\beta\mu$ for different values of the chemical-potential asymmetry~$\beta h$ and compare it to the corresponding results obtained from the third-order virial expansion. Note, in particular, that at the largest values of $\beta\mu$ the relative magnetization is almost $25\%$ for $\beta h = 2.0$.

\subsection{Pressure}
From the density EOS presented in the main text, we are also able to extract the pressure as a function of chemical potential via integrating the Gibbs-Duhem relation:
\beq
  P = \f{\lambda_T^3}{\beta} \int_{-\infty}^{\beta\mu} d(\beta\mu') n(\beta\mu').
\eeq
To perform the integration with minimal numerical errors, we interpolated our data and subsequently integrated the obtained interpolation function, as further discussed below.

Our results in Fig.~\ref{Fig:Pressure} are shown as a function of $\beta\mu$ and for various values of $\beta h$. We observe excellent agreement with the results from the virial expansion at third order across all polarizations studied. Additionally, the experimental data for the balanced case agrees well with our results across a wide temperature range.

\subsection{Finite-volume effects}
As mentioned in the main text, in order to evaluate the path integrals for the UFG, we discretized spacetime using a $(3\!+\!1)$-dimensional lattice. For the spatial volume $V = L^3$ with $L = \ell N_x$, we considered $N_x = 7,\,9,\, 11$ with lattice spacing $\ell = 1$ and periodic boundary conditions. For the temporal direction, we chose $N_\tau = 160$, with temporal lattice spacing $\tau = 0.05 \ell^2$, and anti-periodic boundary conditions for the fermion fields.
Our choice $\beta = \tau N_\tau = 8.0$ determines the thermal de Broglie wavelength $\lambda_T = \sqrt{2\pi \beta}\simeq 7.0$ which is consistent with the continuum-limit window $1 = \ell \ll (\lambda_T, \lambda_F) \ll L = N_x \ell$, where $\lambda_F$ is the Fermi wavelength associated with the interparticle spacing. Thus, the challenge of addressing the systematic effects is that of opening this window of scales by making $N_x$ and $\beta$ as large as possible, in that order, and staying in a sufficiently dilute regime to suppress artifacts associated with the ultraviolet energy cutoff imposed by the lattice.
\begin{figure}[t]
  \centering
  \includegraphics[width=\columnwidth]{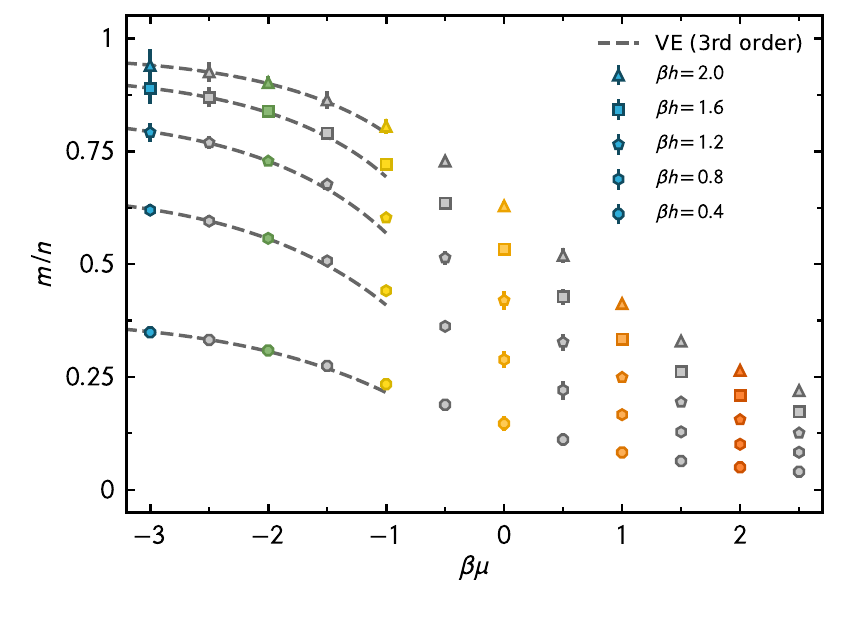}
  \caption{\label{Fig:RelativePol} Magnetization relative to the interacting density as a function of~$\beta\mu$ for various values of~$\beta h = 0.0, 0.4, 0.8, 1.2, 1.6, 2.0$ (symbols from bottom to top). Dashed lines show the corresponding result from the virial expansion evaluated at third order.}
\end{figure}
\begin{figure*}[t]
  \centering
  \includegraphics[width=2\columnwidth]{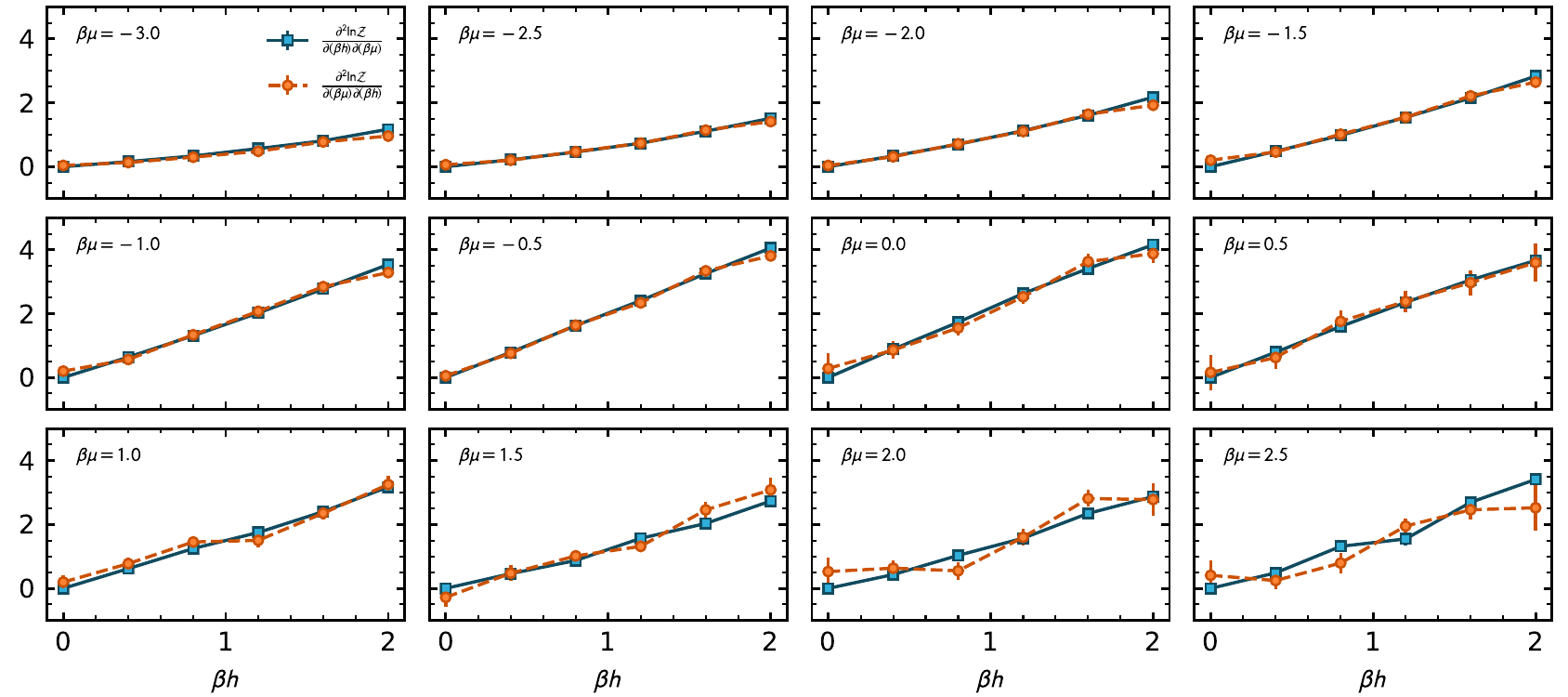}
  \caption{\label{Fig:cross_check} Second derivatives of $\ln\mathcal{Z}$ with respect to $\beta\mu$ and $\beta h$ (blue squares) and with respect to $\beta h$ and $\beta\mu$ (red circles) as a function of $\beta h$ for values of $\beta\mu = -3.0, -2.5, -2.0, \dots, 2.5$. Solid and dashed lines are introduced to guide the eye.}
\end{figure*}
In Fig.~\ref{Fig:volumes}, we display the behavior of the density EOS, in units of $\lambda_T^3$, for a selected set of parameter values~$(\beta\mu,\beta h)$ as a function of inverse spatial volume. Our CL results for the interacting density exhibits a behavior which follows closely the trend of the density of the noninteracting system on an appropriately sized lattice. Moreover, where applicable, our results also show very good agreement with the results from the virial expansion. The scaling of the noninteracting system suggests that our largest lattice sizes are already close to the infinite limit.
\begin{figure}[t]
  \centering
  \includegraphics[width=\columnwidth]{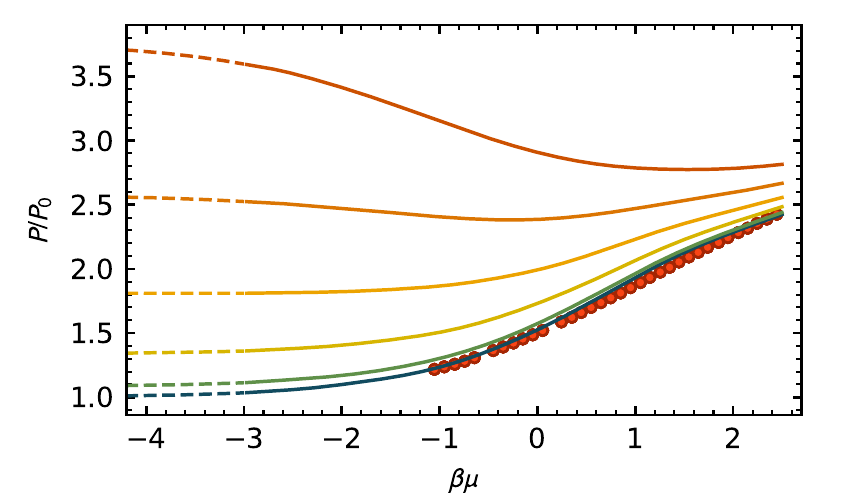}
  \caption{\label{Fig:Pressure} Pressure in units of the noninteracting pressure at corresponding temperature for various values of the chemical-potential asymmetry $\beta h = 0.0, 0.4, 0.8, 1.2, 1.6, 2.0$ (solid lines from bottom to top). Dashed lines show the appropriate third-order virial expansion results and red circles depict experimental values for the balanced gas~\cite{VanHoucke2012}.}
\end{figure}
\begin{figure*}
  \centering
  \includegraphics[width=2\columnwidth]{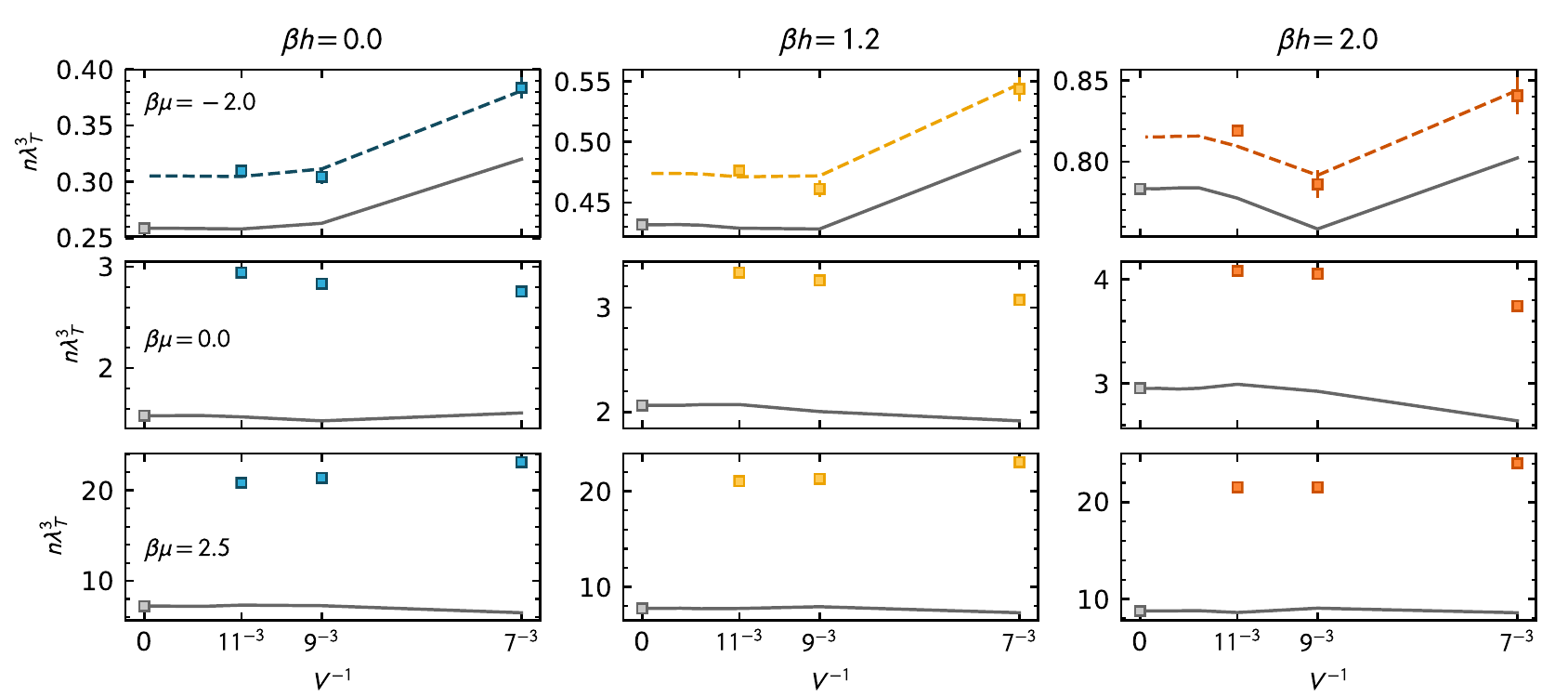}
  \caption{\label{Fig:volumes} Dimensionless density $n\lambda_T^3$ as a function of inverse volume {for selected values of} $(\beta\mu,\beta h)$ together with the noninteracting values on the lattice (solid lines) and the values in the infinite-volume limit~(symbols at $V^{-1} = 0$) as well as, where applicable, the results from the virial expansion (dashed lines).}
\end{figure*}
\begin{figure*}
  \centering
  \includegraphics[width=2\columnwidth]{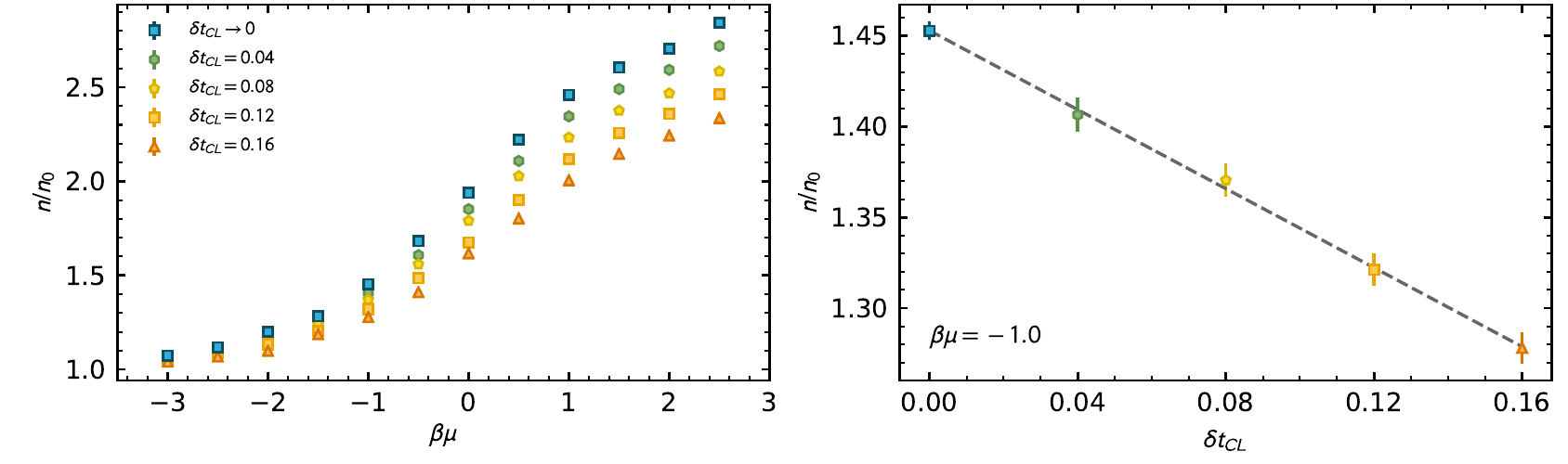}
  \caption{\label{Fig:tcl_extrap} (Left) Density EOS of the balanced gas for values of the CL integration step $\dtcl = 0.16, 0.12, 0.08$ and $0.04$ together with values extrapolated to $\dtcl\rightarrow 0$. (Right) Density EOS for $\beta\mu = -1.0$ as a function of $\dtcl \rightarrow 0$.}
\end{figure*}
%

\subsection{Langevin step-size effects}
Apart from finite-volume effects, Langevin-type approaches suffer from the use of a finite step size $\dtcl$ (required to solve the CL equations). Here, we show the dependence of our results on this parameter, as well as the extrapolation to vanishing step size. {The left panel of Fig.~\ref{Fig:tcl_extrap} shows} the density EOS for various values of the adaptive integration step~$\delta t_{\text{CL}}$, together with the extrapolated values. We observe that the results for the EOS depend on the step size, and the dependence is stronger the larger $\beta\mu$ is. However, the changes always display a linear behavior as $\dtcl \rightarrow 0$ and are therefore well under control.
As an example, this linear behavior is illustrated in detail for~$\beta\mu= -1.0$ in the {right panel of Fig.~\ref{Fig:tcl_extrap}.} Indeed, we in general observe an almost perfect linear behavior which allows for a precise extrapolation to the limit $\dtcl \rightarrow 0$ and thus for an elimination of the systematic error associated with a finite integration step.

\subsection{Interpolations}
In order to obtain static response functions such as the compressibility and magnetic susceptibility, starting from discrete MC data, we carried out interpolations to mitigate numerical errors associated with differentiation and integration. More specifically, the smoothness of our numerical data allowed us to perform a cubic spline interpolation with natural boundary conditions (i.e. the second derivatives on the outermost datpoints are set to zero). Additionally, by solving for the a priori unknown coefficients of the cubic spline, we were able to differentiate and integrate our results exactly on each sub-interval and thus mitigate numerical artifacts. Furthermore, it is possible to propagate the statistical uncertainties through this procedure, once the coefficients are known.

\subsection{Self-consistency checks}
In addition to our analysis of systematic errors, we checked that our results for the density and magnetization are consistent, in the sense that they fulfill the following Maxwell relation:
\beq
\f{\partial^2\ln \mathcal{Z}}{\partial(\beta\mu)\partial(\beta h)} = \f{\partial^2\ln \mathcal{Z}}{\partial(\beta h)\partial(\beta\mu)},
\eeq
The results of this analysis are given in Fig.~\ref{Fig:cross_check}, where we show the second derivatives of $\ln\mathcal{Z}$ with respect to $\beta\mu$ and $\beta h$ and vice versa as a function of $\beta h$ (for multiple values of fixed $\beta\mu$). We find good agreement for almost all values of $\beta\mu$ considered except for the largest value at $\beta\mu = 2.5$, i.e. in the vicinity of the superfluid phase transition, which is also where our results for the density EOS {start} to deviate from the experimental values (as discussed in the main text).
%


\end{document}